\documentclass{acm_proc_article-sp}
\usepackage{placeins}
\begin{document}

\title{Map-Reduce Parallelization of Motif Discovery}

%
%
%
%
%

\numberofauthors{1} 
%
\author{
%
%
\alignauthor
Umang Vipul\\
       \affaddr{The Ohio State University}\\
       \email{umangvipul.1@osu.edu}
}    



\maketitle
\begin{abstract}
Motif discovery is one of the most challenging problems in bioinformatics today.  DNA sequence motifs are becoming increasingly important in analysis of gene regulation. Motifs are short, recurring patterns in DNA that have a biological function. For example, they indicate binding sites for Transcription Factors (TFs) and nucleases. There are a number of Motif Discovery algorithms that run sequentially. The sequential nature stops these algorithms from being parallelized. HOMER is one such Motif discovery tool, that we have decided to use to overcome this limitation.
To overcome this limitation, we propose a new methodology for Motif Discovery, using HOMER, that parallelizes the task. Parallelized version can potentially yield better scalability and performance. To achieve this, we have decided to use sub-sampling and the Map Reduce model. 
At each Map node, a sub-sampled version of the input DNA sequences is used as input to HOMER. Subsampling at each map node is performed with different parameters to ensure that no two HOMER instances receive identical inputs. The output of the map phase and the input of the reduce phase is a list of Motifs discovered using the sub-sampled sequences. The reduce phase calculates the mode, most frequent Motifs, and outputs them as the final discovered Motifs. We found marginal speed gains with this model of execution and substantial amount of quality loss in Discovered Motifs.
\end{abstract}

\section{Introduction}
Bioinformatics is a field that combines the intricacies of Biology and Computer Science. To answer biological questions Biologists collaborate with Computer Scientists to come up with  better techniques. One of the most challenging problems in Bioinformatics today is gene expression and regulation. Gene expression is the process by which Genes, which are sequences of nucleotides in an organism's DNA sequence, eventually form a string of amino acids called Protein. It is the Proteins that separate one species from another or within a species, one individual from another. Gene regulation on the other hand is the phenomenon which decides whether a given gene is expressed at all. 

Motifs are short, recurring patterns in DNA that have a biological function. Motifs are the sites where Transcription Factors (TFs) bind to the DNA sequence and cause gene regulation and expression. We use a tool called HOMER that discovers 'de novo' motifs and returns a list of candidate motifs based on their p-values. The algorithm used by HOMER is innately sequential and thereby limits its scalability. We propose a technique to parallelize the computation performed by HOMER.

The input to HOMER is a number of DNA sequences belonging to an organism. Normally, this is executed in a single machine. We propose sub-sampling the input sequences and using them as input to multiple instances of HOMER running in parallel.  

\section{Design}
The programs were executed in the two instances on the Chris cloud instance. In line with any map-reduce system, there was a master node and a worker node. For Map-reduce, we decided to use bash-reduce and not the well known Apache Hadoop. Bash-Reduce is a very basic implementation of map-reduce framework over inbuilt linux tools like awk, sort etc. The decision to use bash-reduce was based upon the fact that the input data size for motif discovery is not large enough. Hadoop, though very useful with huge data sets, would only have added extra overhead in our case.

Bash-reduce is built upon the standard linux utilities and provides a light map-reduce framework fit for smaller data sets. Without an underlying Distributed File System, files have to be managed separately through linux scripts. All input files are copied to every instance and the output files from each instance is sent to the master node for the reduce function. 
\begin{figure}
\centering
\includegraphics[scale=0.35]{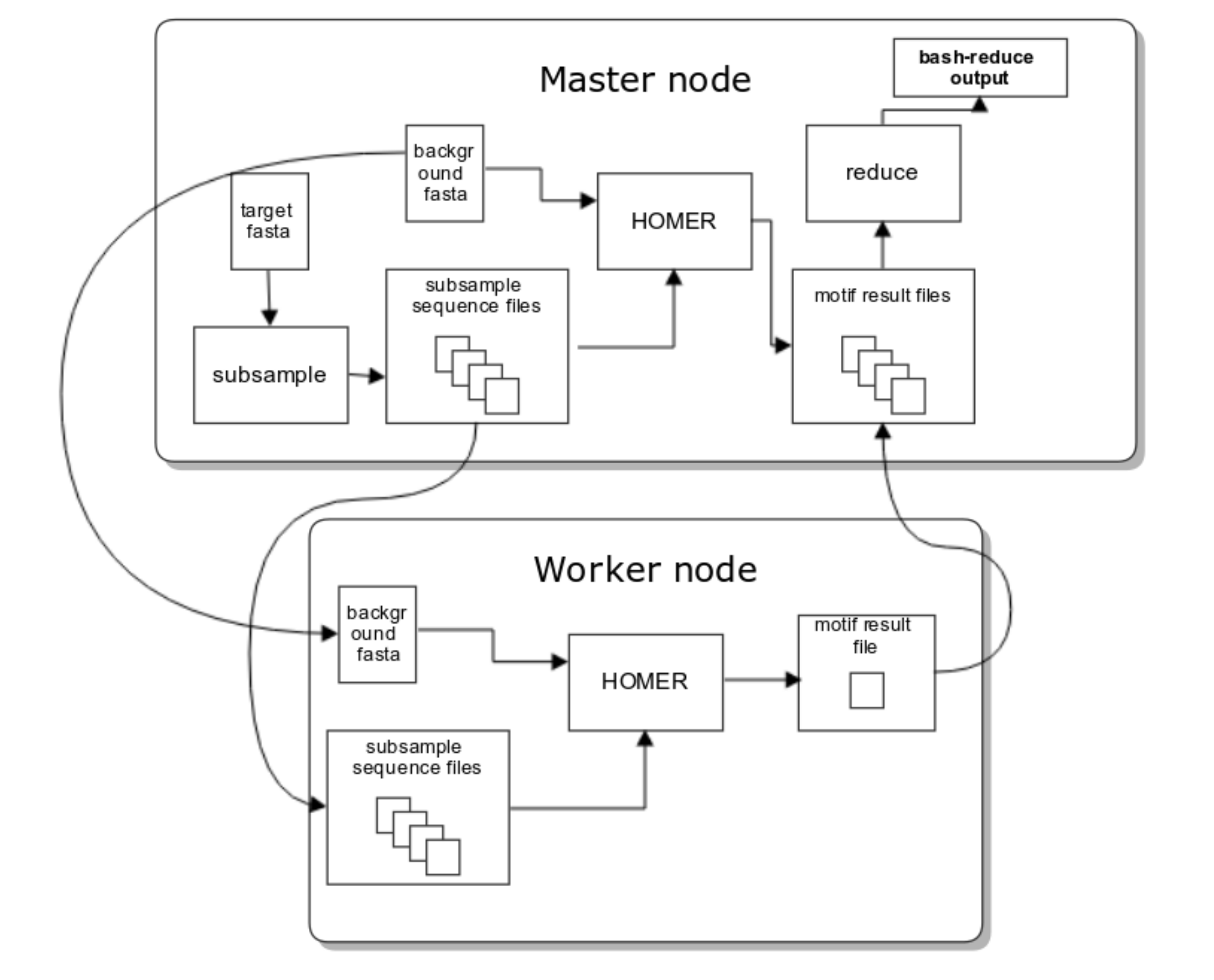} 
\caption{The architecture of map-reduce HOMER}
\end{figure}

\section{Implementation}
The installation of HOMER is pretty simple and is clearly described on its website. An issue that was faced during installation was that directories \textit{cpp} and \textit{bin} had to be created manually before executing the \textit{configureHomer.pl} script. Another necessary package is \textit{weblogo} which is used by HOMER to generate motif logo. It is necessary to add HOMER \textit{bin} path in the \textit{.bash\_profile} file for successful execution of bash-reduce HOMER.

In Bashreduce, the only change was using \textit{nc6} instead of \textit{nc} because of version mismatch of \textit{nc} expected in bash-reduce and the one installed on the instance. 

The python version was upgraded to 2.7.6 because the sample and reduce scripts used function not available in the version installed on the instance. 

There are multiple stages in the execution of HOMER with bash-reduce. Identical to execution in a single machine, we start off with a list of DNA sequences and end with a list of candidate Motifs, but there are multiple steps in between. All the steps mentioned here are carried out by bash or python scripts.

Given the list of DNA sequences, a python script sub-samples the sequences based on the percentage of sequence to be used and number of sub-samples to produce. The DNA sequences are in the well defined FASTA file format. The FASTA file format is defined as a comment prefixed with a ">" for each sequence and then the sequence spread across multiple lines with fixed number of characters in one line, the last line of the sequence being the only exception. The sub-sampling script ensures that the file format is retained. It randomly selects the starting point of the sub-sample such that the desired length of sequence is always selected. The number of sub-sampled files parameter must equal the number of nodes in the map-reduce set up, so that each nodes gets at least one input file. 

Without a distributed file system, all the sub-sampled files are copied over to every worker node. Bash-reduce randomly selects one of the sub-sampled files at each node as input to HOMER at that node. 

The hosts for bash-reduce are specified in the \textit{/etc/br.hosts} file. With the chris cloud, an instance doesn't have access to all the ports and a suitable contiguous port range ( 2 for every node ) must be available in the instances' \textit{/osuml.virtual.dir/ports.map} file. The port number is configured in the \textit{br} script.

After the execution of HOMER, result files from each node are copied to the master node. At the Master node, a Python reduce script parses through results from all the output files to calculate the mode of results. To check for correctness, the output from reduce is compared with the output of serial HOMER execution.

\section{Results}

\begin{table}[h]
\begin{tabular}{c|c|c|c|}
\cline{2-4}
 & \multicolumn{3}{c|}{\textit{\textbf{Mus\_musculus\_74}}} \\ \hline
\multicolumn{1}{|c|}{\textbf{Subsample perc}} & \textbf{Total} & \textbf{node1} & \textbf{node2} \\ \hline
\multicolumn{1}{|c|}{\textbf{25}} & 75.1 & 67.7 & 75.1 \\ \hline
\multicolumn{1}{|c|}{\textbf{55}} & 79.5 & 69.4 & 79.5 \\ \hline
\multicolumn{1}{|c|}{\textbf{75}} & 79.8 & 73.6 & 79.8 \\ \hline
\multicolumn{1}{|c|}{\textbf{90}} & 80.4 & 73.8 & 80.4 \\ \hline
\multicolumn{1}{|c|}{\textbf{100}} & 80.6 & 72.6 & 80.6 \\ \hline
\multicolumn{1}{|c|}{\textbf{Sequential}} & 85.4 & \multicolumn{1}{l|}{} & \multicolumn{1}{l|}{} \\ \hline
\end{tabular}
\caption{Runtimes for \textit{Mus\_musculus\_74} sequence on two nodes}
\end{table}

\begin{table}[h]
\begin{tabular}{c|c|c|c|}
\cline{2-4}
\textbf{} & \multicolumn{3}{c|}{\textit{\textbf{Homo\_sapiens\_75}}} \\ \hline
\multicolumn{1}{|c|}{\textbf{Subsample perc}} & \textbf{Total} & \textbf{node1} & \textbf{node2} \\ \hline
\multicolumn{1}{|c|}{\textbf{25}} & 86.5 & 80.1 & 86.5 \\ \hline
\multicolumn{1}{|c|}{\textbf{55}} & 88.8 & 88.8 & 88.3 \\ \hline
\multicolumn{1}{|c|}{\textbf{75}} & 85.8 & 71.5 & 85.8 \\ \hline
\multicolumn{1}{|c|}{\textbf{90}} & 84.9 & 69.1 & 84.9 \\ \hline
\multicolumn{1}{|c|}{\textbf{100}} & 95.3 & 80.9 & 95.3 \\ \hline
\multicolumn{1}{|c|}{\textbf{Sequential}} & 94 & \multicolumn{1}{l|}{} & \multicolumn{1}{l|}{} \\ \hline
\end{tabular}
\caption{Runtimes for \textit{Homo\_sapiens\_75} sequence on two nodes}
\end{table}

The run times for executing HOMER with our bash-reduce set up are in Table 1 and Table 2. All time values are obtained from average of two runs for that parameter. DNA sequence data for HOMER input was obtained from the Ensembl project. The 2 DNA sequences used were subsets of  \textit{Homo\_sapiens\-.GRCh37.75.dna} and  \textit{Mus\_musculus\-.GRCm38.74.dna}. These sequences functioned as the Target sequence for HOMER. The background sequence for Motif search were created using the \textit{scrambleBG.pl} utility in HOMER. The same background file was used for a given set and all of its various execution parameters.

As tables 1 and 2 show, no substantial reduction is execution times can be observed in the recorded values. The experiments were conducted such that HOMER was the only active process in the two instances. 

Table 5 and Table 6 compare the result obtained from normal HOMER execution and bash-reduce HOMER with various sub-sampling values for both the input sequences used. The results were compared using a bash script. As can be seen, since sub-sampling uses a random number to start the sub--sample sequence, the results vary for each run. Number of motifs found in sub-sampled sequence is always less than motifs in the full sequence. However, not all the motifs found in sub-sampled sequence were found in the original sequence. For every sub-sample, there were motifs found that were not found in original sequence and vice versa.

With Motif discovery, the top results are the most significant, table 3 and table 4 compare quality of results from \textit{br-Homer} in this regard. The top motifs are those that have the lowest log-p value.  These tables count the number of top motifs, discovered from sequential execution, discovered in each execution of \textit{br-Homer}. The values show that the results obtained for the sequence \textit{Mus\_musculus\_74} have more occurrences of top motifs compared to those from \textit{Homo\_sapiens\_75}.

\begin{table*}[h]
\centering
\begin{tabular}{c|c|c|c|c|c|c|}
\cline{2-7}
\textbf{} & \multicolumn{6}{c|}{\textit{\textbf{Mus\_musculus\_74}}} \\ \cline{2-7} 
\textbf{} & \multicolumn{3}{c|}{\textbf{Run 1}} & \multicolumn{3}{c|}{\textbf{Run 2}} \\ \hline
\multicolumn{1}{|c|}{\textbf{Subsample perc}} & \textbf{Top 3} & \textbf{Top 5} & \textbf{Top 10} & \textbf{Top 3} & \textbf{Top 5} & \textbf{Top 10} \\ \hline
\multicolumn{1}{|c|}{\textbf{25}} & 1 & 2 & 3 & 2 & 3 & 5 \\ \hline
\multicolumn{1}{|c|}{\textbf{55}} & 1 & 1 & 3 & 2 & 2 & 4 \\ \hline
\multicolumn{1}{|c|}{\textbf{75}} & 2 & 3 & 5 & 1 & 2 & 5 \\ \hline
\multicolumn{1}{|c|}{\textbf{90}} & 3 & 4 & 6 & 3 & 4 & 6 \\ \hline
\multicolumn{1}{|c|}{\textbf{100}} & 3 & 5 & 10 & 3 & 5 & 10 \\ \hline
\end{tabular}
\caption{Number of common top motifs between sequential and \textit{br-Homer} executions for \textit{Mus\_musculus\_74}}
\end{table*}

\begin{table*}[h]
\centering
\begin{tabular}{c|c|c|c|c|c|c|}
\cline{2-7}
\textbf{} & \multicolumn{6}{c|}{\textit{\textbf{Homo\_sapiens\_75}}} \\ \cline{2-7} 
\textbf{} & \multicolumn{3}{c|}{\textbf{Run 1}} & \multicolumn{3}{c|}{\textbf{Run 2}} \\ \hline
\multicolumn{1}{|c|}{\textbf{Subsample perc}} & \textbf{Top 3} & \textbf{Top 5} & \textbf{Top 10} & \textbf{Top 3} & \textbf{Top 5} & \textbf{Top 10} \\ \hline
\multicolumn{1}{|c|}{\textbf{25}} & 0 & 0 & 0 & 0 & 0 & 0 \\ \hline
\multicolumn{1}{|c|}{\textbf{55}} & 0 & 0 & 2 & 0 & 1 & 1 \\ \hline
\multicolumn{1}{|c|}{\textbf{75}} & 1 & 1 & 2 & 1 & 1 & 3 \\ \hline
\multicolumn{1}{|c|}{\textbf{90}} & 0 & 0 & 0 & 0 & 0 & 1 \\ \hline
\multicolumn{1}{|c|}{\textbf{100}} & 3 & 5 & 10 & 3 & 5 & 10 \\ \hline
\end{tabular}
\caption{Number of common top motifs between sequential and \textit{br-Homer} executions for \textit{Homo\_sapiens\_75}}
\end{table*}

\begin{table*}[h!]
\centering
\begin{tabular}{|c|c|c|c|c|c|c|c|c|l}
\cline{1-9}
 & \multicolumn{4}{|c}{\textbf{Run 1}} & \multicolumn{4}{|c|}{\textbf{Run 2}} &  \\ \cline{1-9}
\multicolumn{1}{|p{1.7cm}|}{\centering \textbf{Subsample perc}}
& \multicolumn{1}{|p{1.30cm}|}{\centering \textbf{br- HOMER}}
& \multicolumn{1}{|p{1.30cm}|}{\centering \textbf{common}}
& \multicolumn{1}{|p{1.30cm}|}{\centering \textbf{HOMER only}}
& \multicolumn{1}{|p{1.30cm}|}{\centering \textbf{br-HOMER only}}
& \multicolumn{1}{|p{1.30cm}|}{\centering \textbf{br-HOMER}}
& \multicolumn{1}{|p{1.30cm}|}{\centering \textbf{common}}
& \multicolumn{1}{|p{1.30cm}|}{\centering \textbf{HOMER only}}
& \multicolumn{1}{|p{1.30cm}|}{\centering \textbf{br-HOMER only}}
 \\ \cline{1-9}
\textbf{25} & 25 & 7 & 68 & 18 & 24 & 15 & 60 & 9 &  \\ \cline{1-9}
\textbf{55} & 27 & 15 & 60 & 12 & 31 & 14 & 61 & 17 &  \\ \cline{1-9}
\textbf{75} & 40 & 23 & 52 & 17 & 42 & 25 & 50 & 17 &  \\ \cline{1-9}
\textbf{90} & 55 & 28 & 47 & 27 & 55 & 28 & 47 & 27 &  \\ \cline{1-9}
\textbf{100} & 75 & 75 & 0 & 0 & 75 & 75 & 0 & 0 &  \\ \cline{1-9}
\end{tabular}
\caption{Motif result quality for \textit{Mus\_musculus\-.GRCm38.74.dna}}
\end{table*}

\begin{table*}[h!]
\begin{tabular}{|c|c|c|c|c|c|c|c|c|l}
\cline{1-9}
\centering
 & \multicolumn{4}{|c}{\textbf{Run 1}} & \multicolumn{4}{|c|}{\textbf{Run 2}} &  \\ \cline{1-9}
\multicolumn{1}{|p{1.7cm}|}{\centering \textbf{Subsample perc}}
& \multicolumn{1}{|p{1.30cm}|}{\centering \textbf{br- HOMER}}
& \multicolumn{1}{|p{1.30cm}|}{\centering \textbf{common}}
& \multicolumn{1}{|p{1.30cm}|}{\centering \textbf{HOMER only}}
& \multicolumn{1}{|p{1.30cm}|}{\centering \textbf{br-HOMER only}}
& \multicolumn{1}{|p{1.30cm}|}{\centering \textbf{br-HOMER}}
& \multicolumn{1}{|p{1.30cm}|}{\centering \textbf{common}}
& \multicolumn{1}{|p{1.30cm}|}{\centering \textbf{HOMER only}}
& \multicolumn{1}{|p{1.30cm}|}{\centering \textbf{br-HOMER only}}  \\ \cline{1-9}
\textbf{25} & 18 & 4 & 71 & 14 & 10 & 0 & 75 & 10 &  \\ \cline{1-9}
\textbf{55} & 26 & 16 & 59 & 10 & 24 & 9 & 66 & 15 &  \\ \cline{1-9}
\textbf{75} & 32 & 20 & 55 & 12 & 41 & 27 & 48 & 14 &  \\ \cline{1-9}
\textbf{90} & 43 & 28 & 47 & 15 & 47 & 31 & 44 & 16 &  \\ \cline{1-9}
\textbf{100} & 75 & 75 & 0 & 0 & 75 & 75 & 0 & 0 &  \\ \cline{1-9}
\end{tabular}
\caption{Motif result quality for \textit{Homo\_sapiens\-.GRCh37.75.dna} }
\end{table*}

\section{Conclusions}
The execution of HOMER to discover Motifs in map-reduce provides marginal drop in execution time with substantial drop in result quality. The results obtained from sub-sampled sequences were inconsistent and incomplete. This clearly outweighs the speed gain and thus makes the whole process questionable.

The marginal speed gains can be attributed to Amdahl's law. The motif discovery algorithm in HOMER consists of multiple steps and only some of them depend on the length of the input sequences. Thus, only those steps speed up but the rest take a similar amount of time eventually resulting in only minor speed gains.

The background sequence used in our experiments were artificially created using the input target sequence. Using actual target and background sequences could have possibly resulted in better result quality.

These results clash with our initial aim of running HOMER in parallel using map-reduce to obtain better performance and scalability. While the proposed model can be scaled well to a certain extent, the gains in execution speed are overshadowed by the loss of result quality.
%
\bibliographystyle{abbrv}
\bibliography{sigproc}  
%
%
\appendix

\section{References}
\begin{enumerate}
\item{HOMER Software - \\http://homer.salk.edu/homer/download.html}
\item{HOMER Motif discovery  tutorial - \\http://homer.salk.edu/homer/motif/index.html}
\item{Seqlogo program - http://weblogo.berkeley.edu/}
\item{Homo Sapiens DNA sequence - \\http://www.ensembl.org/Homo\_sapiens/Info/Index}
\item{Mus musculus DNA sequence - \\http://useast.ensembl.org/Mus\_musculus/Info/Index}

\item{D'haeseleer, Patrik: What are DNA sequence motifs?, Nature Biotech, Vol 24 Issue 4, April 2006}

\item{Heinz S, Benner C, Spann N, Bertolino E et al. Simple Combinations of Lineage-Determining Transcription Factors Prime cis-Regulatory Elements Required for Macrophage and B Cell Identities. Mol Cell 2010 May 28;38(4):576-589. PMID: 20513432}

\item{R. Crowley. BashReduce - Crowley Code.\\ https://github.com/erikfrey/bashreduce.}

\item{Sundeep Kambhampati, Jaimie Kelley, Christopher Stewart, William C.L. Stewart, Rajiv Ramnath. "Managing Tiny Tasks for Data-Parallel, Subsampling Workloads"}

\item{Huang, Hailiang, Sandeep Tata, and Robert J. Prill. "BlueSNP: R package for highly scalable genome-wide association studies using Hadoop clusters." Bioinformatics 29.1 (2013): 135-136.}

\item{Richard Sampson, Ming Yang, Siyuan Wei , Chaitali Chakrabarti, and Thomas F. Wenisch. "Sonic Millip3De:A Massively Parallel 3D-Stacked Accelerator for 3D Ultrasound"}

\end{enumerate}


\balancecolumns
\end{document}